# Moulding hydrodynamic 2D-crystals upon parametric Faraday waves in shear-functionalized water surfaces.

Mikheil Kharbedia[1], Niccolò Caselli[1], Horacio López-Menéndez[1], Eduardo Enciso[1] and Francisco Monroy[1,2,*]

[1] Department of Physical Chemistry, Universidad Complutense de Madrid, Ciudad Universitaria s/n, E28040 Madrid (Spain).
[2] Translational Biophysics, Instituto de Investigación Sanitaria Hospital Doce de Octubre, E28041 Madrid (Spain).

## Abstract

Faraday waves (FWs), or surface waves oscillating at half of the natural frequency when a liquid is vertically vibrated, are archetypes of ordering transitions on liquid surfaces. The existence of unbounded FW-patterns sustained upon bulk frictional stresses has been evidenced in highly viscous fluids. However, the role of surface rigidity has not been investigated so far. Here, we demonstrate that dynamically frozen FWs – that we call 2D-hydrodynamic crystals – do appear as ordered patterns of nonlinear surface modes in water surfaces functionalized with soluble (bio)surfactants endowing in-plane shear stiffness. The strong phase coherence in conjunction with the increased surface rigidity bears the FW-ordering transition, upon which the hydrodynamic crystals were reversibly moulded by parametric control of their degree of order. Crystal symmetry and unit cell size were tuned depending on the FW-dispersion regime. The hydrodynamic crystals here discovered could be exploited in touchless strategies of soft matter scaffolding. Particularly, the surface-directed synthesis of structured materials based on colloids or polymers and cell culture patterns for tissue engineering could be ameliorated under external control of FW-coherence.

* Corresponding email: monroy@ucm.es

## Introduction

Nonlinear surface waves (NLSWs) in fluids embrace a richness of hydrodynamic behaviors dated back to Faraday, who discovered the existence of bounded oscillations at the free surface of vibrated liquids[1]. The Faraday waves (FWs) are resonant oscillations of the unbalanced liquid weight restored by surface tension, being characterized by a subharmonic response (parametrically resonant mode with frequency one half of the forcing excitation)[2]. A critical driving force defines the onset for the Faraday instability[3], which has been prescribed as a hydrodynamic regime for NLSW-ordering in the macroscopic realm[4,5]. The existence of unbounded NLSW patterns (stripes, square or triangular unit cell, and even quasi-crystals) at large aspect-ratio have been reported in highly viscous fluids close to the Faraday instability[6–9], where surface ordering is scaffolded upon bulk frictional stresses[9,10]. That bulky class of FW-patterns have been exploited, for instance, to organize granular layers[11], achieve colloidal lattices on top of structured suspensions[12], or in templating cell culture patterns[13,14]. Moreover, container-bounded FWs have been employed to create standing patterns in inviscid liquids[15,16], in analogy with the parametric excitations of quantum condensates[17,18] and the optical lattices able to confine ultra-cold atom gases[19]. As an innovative synthesis concept in materials science, externally guided FW-patterning is modernly envisaged as a powerful liquid-based templating approach for material micromoulding.

In inviscid liquids, free-standing NLSWs are identified as phase-randomized turbulent cascades of harmonics excited upon resonant coupling with a pumping mode[20]. These disorganized NLSWs excited below the FW-threshold (i.e. without subharmonic response), are well-known in the capillary regime of weak turbulence in which they display a characteristic Kolmogorov-Zakharov (KZ) spectrum[22]. Higher amplitudes lead to FW-excitation[23], even stronger NLSW turbulences[23], which have been studied in the context of hydrodynamic structuring built upon dissipative vortices[24]. Arguably, intrinsic surface elasticity could harness the turbulent randomness into coherence domains able to sustain NLSW-ordering. Despite the enormous avenues of applications that can be envisioned upon wave freezing in liquid surfaces, the generation of stiffness-induced NLSW-patterns has been not explored so far.

In this paper, by addressing in detail the NLSW-regimes driven on water surfaces, we focus on the coherent states that could entail surface wave order upon in-plane shear stiffening endorsed by an adsorption film. By taking advantage of the archetypal Faraday experiment, we systematically analyze the parametric FW-regime by using an optical probe that relied on laser Doppler vibrometry (LDV). We demonstrate the emergence of stationary FW-patterns in the rigidity-functionalized surfaces under no influence of wave bounding conditions, even in the absence of a bulky frictional skeleton. Since these novel surface patterns resemble a two-dimensional crystalline order, we call them as "hydrodynamic 2D-crystals".

In pursuit of discovering the role of surface stiffness in forming hydrodynamic crystals, we engineered the free surface of water by adsorbing rigid films of soluble surfactants at the air/water (A/W) interface. In particular, we employed β-aescin, which is a saponin biosurfactant (with anti-inflammatory and vasoconstrictor effects[25]), capable of forming rigid monolayers due to hydrogen bonding[26–29]. When FWs are tuned upon external driving, hydrodynamic 2D-crystals can be moulded in terms of the excitation characteristics. Since bulk friction remains as low as corresponds to the inviscid water subphase, we demonstrate reversible surface rigidization as the key-player promoting the mode freezing necessary to form the novel class of hydrodynamic 2D-crystals.

## Results

**NLSWs excited on water surfaces.** To implement a NLSW-platform for hydrodynamic templating, we exploited gravity-capillary waves (GCWs) parametrically excited on liquid surfaces (see Methods). Figure 1a depicts the experimental setup used to drive and probe nonlinear GCWs upon periodic monochromatic excitation. The forcing device was designed to vertically vibrate a liquid container of large lateral dimensions (cylindrical diameter $D = 20\ cm$) at a variable amplitude ($A$) and fixed driving frequency ($\omega_0$), which determine the driving acceleration as $a = A\omega_0^2$ (see Methods). As hydrodynamic paradigm, we considered an incompressible Newtonian fluid (water) with a density ($\rho$) and surface tension ($\sigma$) adequate to support nonlinear GCWs[30]. We performed experiments spanning across the critical Faraday acceleration[31] ($a_F$), which tags the onset of the parametric resonance exhibiting subharmonic response at $\omega_0/2 \equiv \omega_{1/2}$. By focusing on the capillary domain ($\omega_0 \geq \omega_c$), the FW-threshold is given in terms of the kinematic viscosity ($\mu$) as[31]:

$$a_F = 8\mu(\rho/\sigma)^{1/3}(2\pi\omega_0)^{5/3}. \tag{1}$$

An extension to the complete GC-domain can be also made by using the effective tension $\sigma_{eff} = \sigma(1 + \rho g \lambda^2/4\pi^2 \sigma)$ defined in terms of the mode wavelength $\lambda$ (see Methods).

This relationship (1) was validated for pure water and the other surfaces considered in this work (see Supplementary Tables T1-T2 and Supplementary Figure S1). As a reference frequency we choose $\omega_0 = 47$ Hz, for which Eq. (1) predicts $a_F \approx 2\ m/s^2$ in water at room temperature ($\sigma = 0.072\ N/m$, $\rho = 10^3\ kg/m^3$ and $\mu \approx 2\ m^2/s$). This arbitrary choice lies in the CW domain ($\omega_0 \gg \omega_c \approx 14$ Hz), where free-standing FWs propagate in the deep-water regime as NLSW-ensembles of large aspect-ratio with wavelengths shorter than the vessel dimensions ($\lambda_0 \approx 6\ mm \ll D$) (Fig 1b; see Methods). Similar results were obtained for GWs at frequencies below (but close) to $\omega_c$ (see Fig. 1b; top panel and Supplementary Table T1). Using this hydrodynamic platform, a wide NLSW-scenario was explored in terms of a reduced driving acceleration $\Gamma \equiv a/g$; hereinafter, the control parameter $\Gamma$ will be referred to the dimensionless Faraday threshold for pure water ($\Gamma_F \equiv a_F/g \approx 0.2$).

Figure 1c displays the waving textures captured in the relevant amplitude domains for increasing driving force. Below a Hookean limit ($\Gamma < \Gamma_0$), despite the low amplitude CWs are invisible to the naked eye (see Fig. 1c; top panel), LDV detected the linear surface response as a monochromatic transversal wave appeared at $\omega_0$ (see Supplementary Figure S2). At moderate driving force ($\Gamma_0 < \Gamma < \Gamma_F$), we observed high-amplitude surface undulations as circular waves mastered by the fundamental wavelength $\lambda_0$ (see Fig. 1c; second panel). Further excitation above the Faraday onset ($\Gamma > \Gamma_F$), caused progressive surface roughening as scrambled ripples of variable amplitude, which were associated to the FW-parametric resonance (see Fig. 1c; third panel). Finally, at very high driving forces ($\Gamma > \Gamma_{chaos} \gg \Gamma_F$), increasing waving disorder gave rise to a dynamically chaotic surface (see Fig. 1c; bottom panel). At $\Gamma \gg \Gamma_{chaos}$, the surface became unstable even observing ejection of droplets (see outset in Fig. 1c; bottom). These amplitude domains were achievable not only with CWs of variable frequency but also with GWs below $\omega_c$ (see Supplementary Figure S3). All these NLSW-states were robustly reproduced in water and in other liquids, however, we found that the transitions between them are strongly dependent on the liquid viscosity (see Supplementary Table T1).

Because the low viscosity of water warrants NLSW propagation at moderate wave damping[30,31], but is not sufficiently high for sustaining free-standing surface wave patterns built upon bulk friction[14], the results displayed in Fig. 1 validate our NLSW-platform for GC-unbounded waves with a parametric control of the wave amplitudes. Furthermore, in the Faraday regime, at $\Gamma > \Gamma_F$, we assured the fulfillment of the large aspect-ratio condition[32], and prevented undesired stationary waves upon edge bounding.

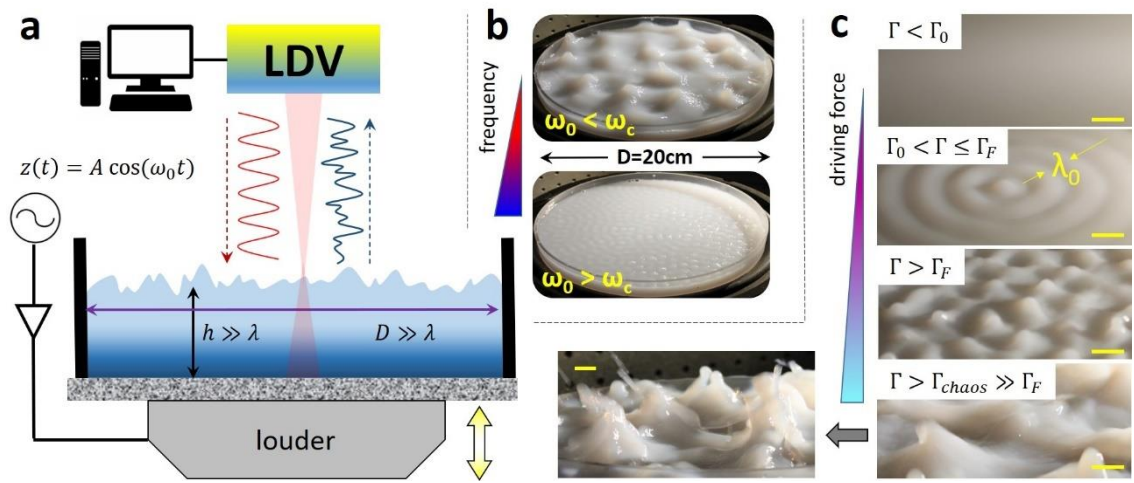

**Figure 1. a)** Schematics of the experimental setup. Surface waves were generated at the A/W-interface under vertical agitation in a cylindrical container with large aspect-ratio (diameter $D = 20$ cm and height $h = 2$ cm, both larger than the typical induced oscillation wavelength $\lambda$). The container was attached to an electromechanical driver (louder) that induced a periodic vertical displacement $z(t)$ at frequency $\omega_0$ and amplitude $A$. The surface wave field displacement was recorded as a function of time by means of a Laser Doppler Vibrometer (LDV). **b)** Images of the investigated surface for water mixed to a low concentration of colloidal particles (PS-MMA) acting as white colorant (see Materials), when the vessel was vertically vibrating in the gravity ($\omega_0 < \omega_C$) and capillary

regime ($\omega_0 > \omega_C$), respectively. **c)** Images of characteristic surface response for increasing driving force, reported in terms of the reduced acceleration $\Gamma \equiv A\omega_0^2/g$, in the capillary wave regime ($\omega_0 = 47$ Hz, hence $\lambda_0 = 5.9$ mm; see Methods). From top to bottom panel: low amplitude ($\Gamma < \Gamma_0$) exhibiting a linear response; intermediate amplitude ($\Gamma_0 < \Gamma < \Gamma_F$) exhibiting concentric waves; amplitude above the Faraday threshold ($\Gamma > \Gamma_F$) displaying surface scrambled ripples; extremely high amplitude ($\Gamma > \Gamma_{chaos} \gg \Gamma_F$) where chaotic waves give rise to droplet ejections (see the bottom left panel with enlarged view). The scale bar is 5 mm.

**NLSW turbulence domains: cascades of pure FWs are coherent and strongly coupled.** Figure 2a shows the spectral features determined by LDV in the NLSW-states spanned by varying $\Gamma$ with respect to $\Gamma_F$. At low driving force ($\Gamma = 0.1 < \Gamma_F$; Fig. 2a, first row), the nonlinear surface response emerged as a cascade of discrete harmonics exhibiting peaks at multiple integers of the fundamental frequency ($\omega_n = n\omega_0$). The recorded cascade followed the Kolmogorov-Zakharov (KZ) intensity decay, i.e. with power spectral density $(PSD) \sim \omega^{-17/6}$, characteristic for weak CW-turbulence[33,34]. A homogenous distribution of the mode phase was observed (see Fig. 2b, first panel), which confirmed the natural phase randomness of the NLSWs created in the KZ-regime at $\Gamma < \Gamma_F$. At the onset of the Faraday instability ($\Gamma = 0.2 \approx \Gamma_F$; Fig. 2a, second row), LDV revealed the emergence of the subharmonic peak ($\omega_{1/2} \equiv \omega_0/2$), in addition to the fundamental mode ($\omega_0$). Moreover, two superposed cascades decay alternating within the lower harmonics up to a Faraday cut-off ($n \lesssim n_F \approx 20$). They corresponded, respectively, to the ordinary KZ-cascade of the fundamental mode ($o \equiv$ KZ at $n\omega_0$), and an extraordinary cascade of the Faraday subharmonic ($e \equiv$ F at $n\omega_{1/2}$). Both cascades showed similar decay rates as $PSD \sim \omega^{-5}$, consistent with cooperative scaling characteristic of strong, highly correlated, wave turbulence[24]. Interestingly, those composite FW-cascades ceased at $n_F$, which corresponded to the highest mode before incoherent KZ-turbulence became dominant. Indeed, for $n \geq n_F$, the capillary-like $\omega^{-17/6}$-decay remerged (see also Supplementary Figure S4). This KZ-Faraday (KZF) hybridization was only detected around $\Gamma \gtrsim \Gamma_F$, where phase-locking was still weak (see Fig. 2b, second panel). Above the Faraday onset ($\Gamma \geq \Gamma_F$; Figs. 2a-b, third row), the composite ($o \cup e$) cascade appeared as a unified FW-cascade decaying as $PSD \sim \omega^{-5}$. Here, two mode-coupling features were detected: i) peak broadening indicating inter-cascade energy exchange ($o \leftrightarrow e$), and ii) phase-locking leading to intermodal coherence. Hereinafter, we will refer to this class of turbulent cascade as pure-FWs (see also Supplementary Figure S5). Far beyond the Faraday threshold ($\Gamma \gg \Gamma_F$; Fig. 2, fourth row), we observed a transition towards a chaotic regime at $\Gamma > \Gamma_{chaos} \approx 0.9$. In this domain, the observed cascades were characterized by a Lorentzian decay as $PSD \sim \omega^{-2}$, peak broadening, and phase decoherence, which evidenced the disordered nature of the chaotic interactions (Brownian-like)[35]. Since the spectra did not show discrete resonances but a frequency superposition (see Supplementary Figure S6), we refer to those as continuous chaotic spectra by reference to the Landau's conjecture for chaotic flows[36].

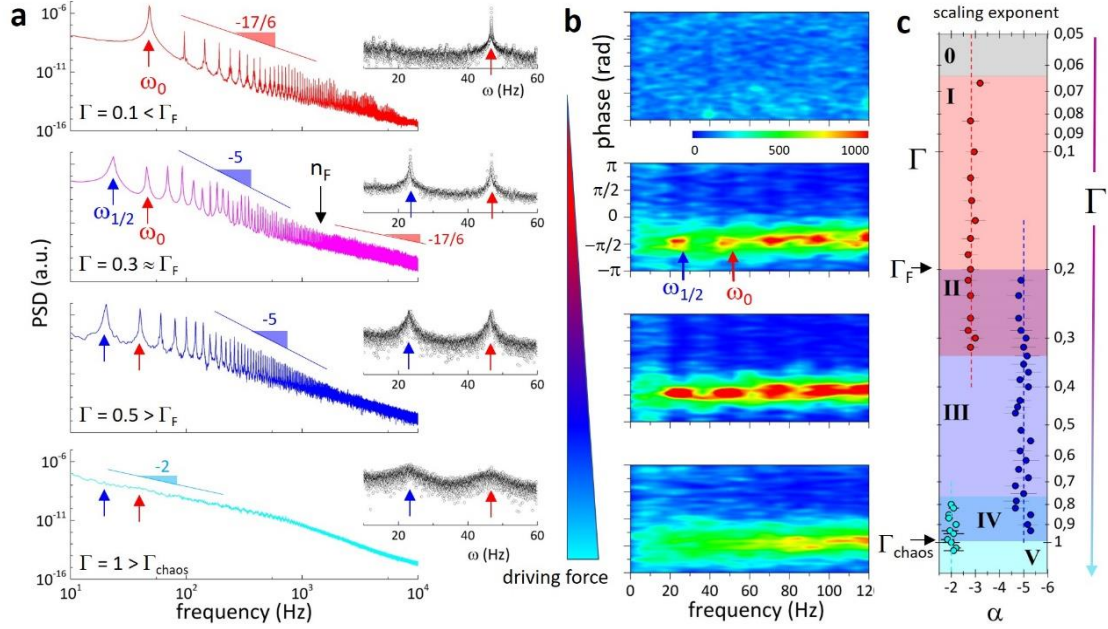

**Figure 2**. **a)** Power spectral density (PSD) of the A/W-interface obtained by vertically oscillating the vessel at frequency $\omega_0 = 47$ Hz for increasing acceleration $\Gamma$ from top to bottom. *i*) Kolmogorov-Zakharov spectrum of weak wave turbulence ($\Gamma < \Gamma_F$) with $\omega^{-17/6}$-intensity decay typical of cascades of nonlinear capillary waves; *ii*) Hybrid spectrum of turbulence with superimposed subharmonic $\omega_{1/2}$ (blue arrow) and fundamental $\omega_0$ (red arrow) cascades ($\Gamma \geq \Gamma_F$). The Faraday ($\omega^{-5}$) cascade predominates in the lower frequency range, while $\omega^{-17/6}$- decay prevails at higher frequencies. *iii*) Pure Faraday spectrum ($\Gamma > \Gamma_F$); *iv*) chaotic continuous spectrum ($\Gamma \gg \Gamma_F$). The insets highlight the spectral range of $\omega_0$ and $\omega_{1/2}$ in linear scale. **b)** Phase distribution of the spectra reported in a). Phase locking between the cascade resonances is observed when $\omega_{1/2}$ emerges. **c)** Scaling exponent ($\alpha$) that fits the peak cascades as a function of $\Gamma$, highlighting the transitions between the regimes described in a); they are qualitatively delimited by different color areas. The gray (0) area represents the linear regime; pink (I) area the $\omega^{-17/6}$-KZ decay; the purple (II) area the hybrid cascade (KZF); the dark blue (III) area the pure Faraday cascade ($\omega^{-5}$); the blue (IV) area the coexistence of Faraday and $\omega^{-2}$-decay cascades; finally, the light blue (V) area delimits the completely chaotic cascades.

To highpoint the transitions between the aforesaid NLSW-states, Figure 2c shows the fitted values of the scaling exponents $PSD \sim \omega^{-\alpha}$ as spanned in terms of six regimes (labelled from 0 to V; see caption). Pure-FWs underlying strong resonant coupling and phase coherence were found in a relatively broad domain of driving forces (state III: $1.5\Gamma_F \lesssim \Gamma \lesssim 4\Gamma_F$). This genuine FWs-domain is flanked by hybrid regimes contaminated by decoherent interactions. Similar sequences of waving states were observed at different driving frequencies either in water or with other liquids (see Supplementary Figure S7), which suggests a universal NLSW-behavior as depicted in the "state diagram" in Figure 2c.

**Surface shear-functionalization with β-aescin induces hydrodynamic crystals upon FW excitation.** After characterizing the NLSW-states, here we will deliver the main outcome of this work: the induction in the coherent FW-state of large hydrodynamic 2D-crystals after inducing surface-shear rigidization. As an efficient stiffening agent, we employed β-aescin reversibly adsorbing as a rigid monolayer at the A/W-interface. The surfactant was dissolved at a concentration close to its critical micellar concentration ($cmc \approx 0.4$ mM). Then, we excited NLSWs by forcing the hydrodynamic states to lie in the pure FW-domain. Figure 3a compares the FW-textures observed under parametric excitation either in a bare water surface (top panel) or in a covered surface functionalized with β-aescin (bottom panel). The occurrence of a coherent *surface wave freezing* induced by the presence of the surfactant is remarkable since the equivalent bare surfaces appeared completely disordered under identical waving excitation. As a matter of fact, only the pure-FWs could arrange into stationary patterns with a long-range order forming a macroscopic 2D-crystal (see Supplementary Figure S8).

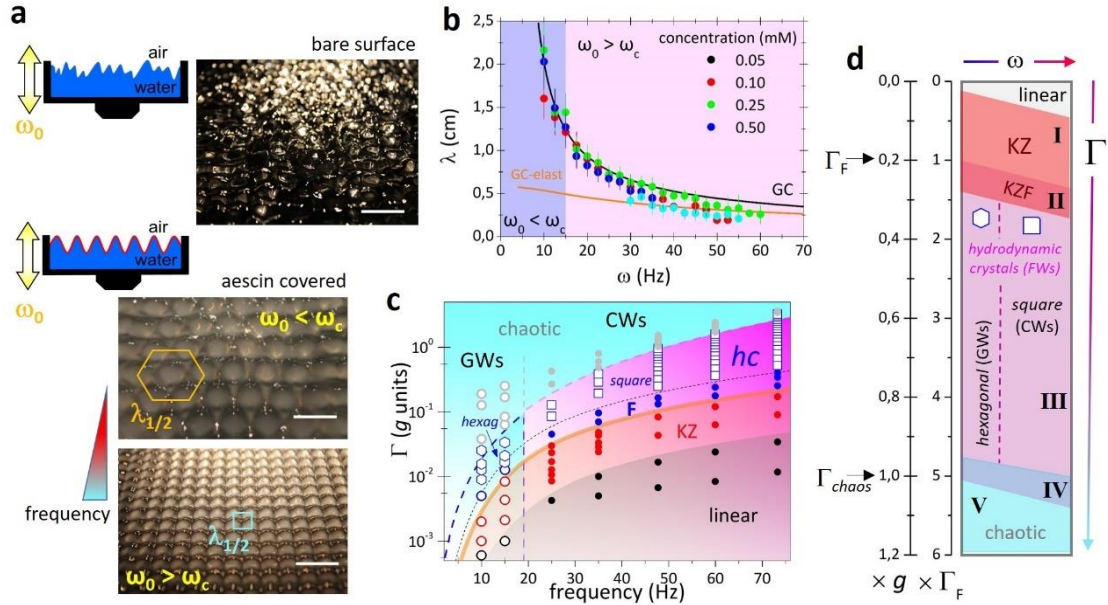

**Figure 3.** Hydrodynamic crystal formation **a)** Schematics and real images of the bare water surface (top) and for the water surface covered with β-aescin (bottom), under the same parametric excitation in the pure Faraday regime. Two distinct arrangements can be tailored: hexagonal packing in the gravity wave regime ($\omega_0 < \omega_c$) and square packing in the capillary wave regime ($\omega_0 > \omega_c$). The ordering is induced by a rigid surface functionalization due to adsorption of soluble β-aescin monolayers (solid-like at $cmc \approx 0.4 \text{ mM}$). **b)** Hydrodynamic crystals dispersion relation measured for different aescin concentration (solid dots), along with theoretical dispersion predictions for pure GC-waves ($\sigma \approx 30 \text{ mN/m}$; black line) and elasticity-controlled GC-waves ($G/\sigma \approx 30$; orange line, see Supplementary Note N2). The wave characteristics correspond to the Faraday subharmonic, i.e. $\omega \equiv \omega_{1/2} = \omega_0/2$ and $\lambda \equiv \lambda_{1/2}$. **c)** State diagram in the $\Gamma - \omega$ space for parametrically NLSWs in water functionalized with aescin. Linear (black dots), Kolmogorov-Zakharov (KZ; red dots), Faraday (F; blue dots) and chaotic (gray dots) spectral cascades are identified, along with the emerging states corresponding to hydrodynamic crystals (hc) and their geometry reported as white hexagons and square, respectively. The distinct regimes are qualitatively delimited by lines and color areas. Note that hydrodynamic crystals occur in the pure Faraday regime slightly above $\Gamma_F$. **d)** Simplified schematics of the state diagram evaluated in c).

Notably, crystal symmetry can be harnessed depending on the dispersion regime (see Fig. 3a, bottom panel). By tuning the driving frequency $\omega_0$ with respect to the capillary frequency $\omega_c$, we obtained either hexagonal crystals with the triangular symmetry in the gravity regime ($\omega_0 < \omega_c$), or square crystals in the capillary regime ($\omega_0 > \omega_c$). The size of the side of the unit cell was fixed by the FW-subharmonic wavelength $\lambda_{1/2} \equiv 2\lambda_0$, which varies with $\omega_{1/2}(\equiv \omega_0/2)$ as determined by the GC-dispersion equation (for given $\sigma$ and $\rho$ under nonlinear corrections due to surface elasticity; see Fig. 3b and Supplementary Note N2). These quantitative results, complemented with the graphical evidence in Supplementary Figure S8, prove that the crystalline structure can be moulded under parametric control (by varying $\Gamma$ at fixed $\omega_0$). Figure 3c reports the state diagram found in water surfaces covered by β-aescin, as determined by the systematic cartography of the $\Gamma - \omega$ space. Figure 3d shows that although FWs exist in a broad interval of driving acceleration ($\Gamma_F < \Gamma < \Gamma_{chaos}$), hydrodynamic crystals with a long-range order were only achieved well inside the pure-FW domain of the covered surface (state III; Fig. 3d) The upper and bottom limits of this domain of frozen patterns remain almost unaltered with respect to its homologous state in the bare surface (see Fig. 2c).

**Coherent Faraday "freezing" occurs upon phase locking.** To scrutinize the hydrodynamic skeleton underlying the 2D-ordering in NLSW-patterns, we identified coherent couplings in the templating NLSW-cascades. The relevant conclusions are emphasized in Figure 4 focusing on CWs, although similar qualitative conclusions were drawn for GWs. As shown in Supplementary Figures S8 and S9, the surfaces covered with β-aescin followed the same sequence of NLSW-states observed in the bare water surface (see Fig. 2c), exhibiting similar spectral decay scaling for all the five regimes (see Fig. 4a). The presence of ordered crystals was uniquely detected in the regime of genuinely coherent FWs (Fig. 4b and Fig. S8). In this case the pure-FWs resonant harmonics were characterized by extreme spectral narrowing (Fig. 4c) and highly localized phase-locking (Fig. 4d). However, phase decoherence elicited resonance narrowing and progressive surface disordering. Even though, some degree of deterministic organization remained in the chaotic V-state at $\Gamma > \Gamma_{chaos}$ (see Fig. 4; bottom panels), where phase coherence persisted higher than in the genuinely disordered I-state at $\Gamma < \Gamma_F$ corresponding to incoherent KZ-turbulence (Fig. 4; top panels).

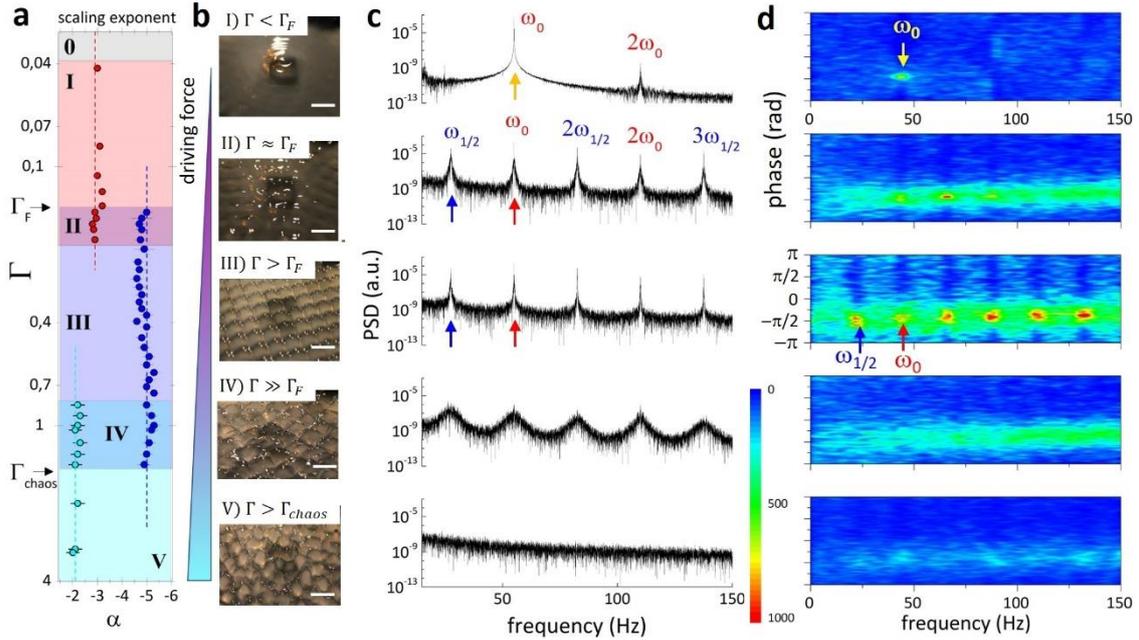

**Figure 4**. Scaling exponent ($\alpha$) obtained from the complete spectral cascades (see Fig. S9), highlighting the transitions between the waving states qualitatively delimitated by different color areas (as in Fig. 2c). **b)** Surface imaging of the NLSW-states corresponding to the subsequent spectral analyses performed at the center of the vessel for vibrating frequency $\omega_0 = 47$ Hz in the CW-regime: **c)** power spectral density (PSD) and **d)** phase distributions at increasing amplitude (spectral zooming at the fundamental modes and first harmonics; from top to bottom): I) Weak KZ-turbulence ($\Gamma < \Gamma_F$) exhibiting only the harmonic cascade, small coherence between the peaks and a disordered surface. II) Hybrid KZF-cascade at the onset of the Faraday regime ($\Gamma \approx \Gamma_F$), where the subharmonic cascade at $\omega_{1/2}$ emerged, the peaks acquired a higher degree of coherence and the surface displayed a waving pattern but still not ordered. III) Pure FWs in the regime $1.5\Gamma_F \lesssim \Gamma \lesssim 4\Gamma_F$, where the peaks narrowed exhibiting a strong coherence and the FWs "frozen" as a stable square CW-pattern. IV) For increasing acceleration ($\Gamma_F \ll \Gamma < \Gamma_{chaos}$) the peaks broadened and decoherence broke up the crystal. V) In the complete chaotic regime ($\Gamma > \Gamma_{chaos}$), NLSW-spectra did not show any peak, the phase distribution was random, and the waving surface became strongly disordered.

**Material relationship.** By exploiting experimental interfacial rheology (see Methods and Supplementary Note N3), we performed a quantitative analysis of the degree of waving ordering by reference to the in-plane shear modulus ($G$). Figure 5a shows the results for different surfactant additives. For β-aescin monolayers, we found $G$ increasing with concentration (up to $G \approx 1\,N/m$ as highest at $c \geq 2cmc$). This significant surface stiffening elicited a progressive intensification of the degree and spatial range of the crystalline order on the hydrodynamic crystal (see Fig. 5b). This paradigmatic stiffener was compared with other surface functionalizations that did not induce enough rigidity to support NLSW-patterning, such as soluble CTAB forming fluid monolayers ($G \approx 0$)[37], and insoluble DPPC at the gel phase of monolayer packing ($G \approx 10^{-4}\,N/m$)[38]. Among the tested surfactants, only β-aescin was able to harness FWs as 2D-hydrodynamic crystals, whereas the other surface monolayers did not support ordered patterns (see Fig. 5c; left panels). The efficiency of β-aescin as 2D-rigid fabric to freeze FWs is remarkable since, under the same conditions, the bulky skeletons scaffolded on the 3D-frictional stresses of colloidal solutions were unable to support hydrodynamic crystals despite their higher dimensionality (see Fig. 5c; right panels).

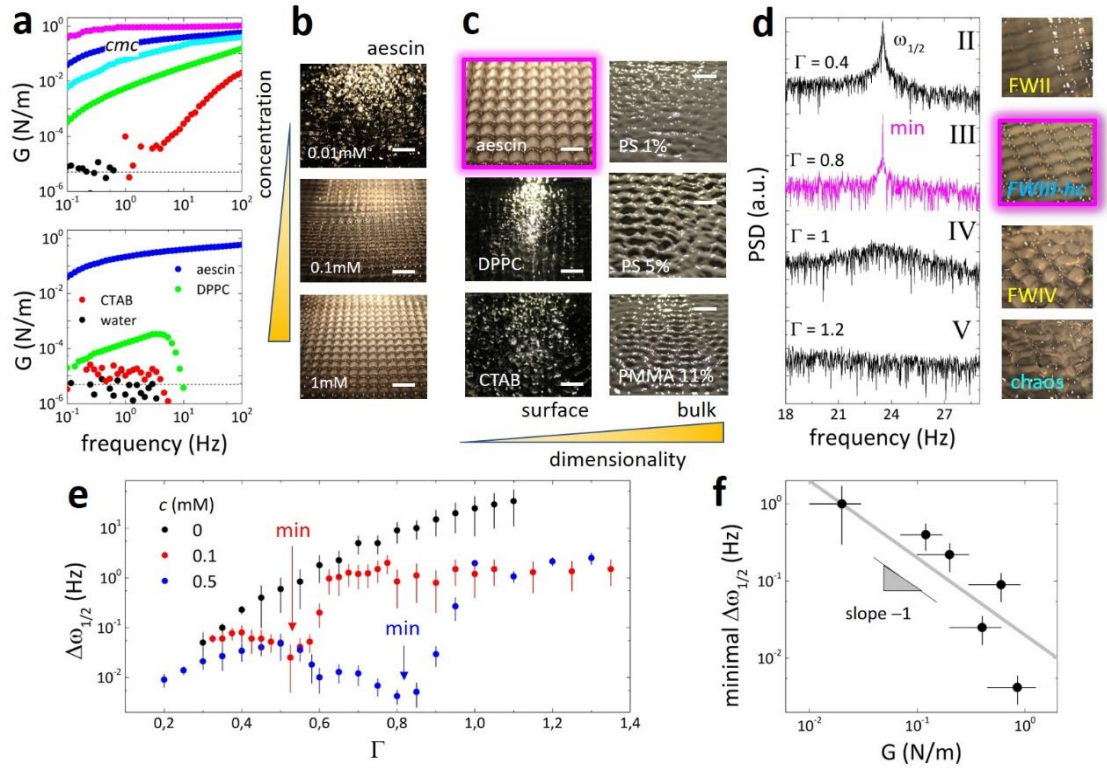

**Figure 5. a)** Linear interfacial rheology (see Supplementary Note N3) of aqueous solutions for different aescin concentration (top panel; concentrations in mM: • 0 (pure water); • 0.01; • 0.04; • 0.1; • 0.4 and • 1) and different surfactants as CTAB and DPPC (lipid) at equivalent concentration as the surface tension of the aescin solution at $cmc = 0.4 mM$ ($\sigma \approx 30\, mN/m$; lower panel). The surface stiffness ($G$) varies by several orders of magnitude from the highest saturation value $G \approx 1\,N/m$ at 1mM concentration ($\approx 2.5 cmc$; magenta), 0.4mM ($\approx cmc$; blue), 0.1mM (cyan), 0.05mM (green), and 0.01mM (red). The black symbols correspond to the bare water surface, which are compatible with the instrumental sensitivity (ca. $5\,\mu N/m$). **b)** Surface waving images at different aescin concentration under driving frequency $\omega_0 = 47$ Hz and acceleration $\Gamma > \Gamma_F$, which allow the formation of square-lattices of hydrodynamic crystals at $G \geq 0.1\,N/m$, corresponding to aescin concentrations $c \geq 0.1 mM \approx 0.25 cmc$. **c)** Surface waving images with aescin at $c \approx cmc$ leading to hydrodynamic patterns, and two surfactant additives (CTAB and DPPC) unable to harness surface ordering at equivalent surface tension ($\sigma \approx 30\,mN/m$; see main text for details). The right panels compare the disordered waves obtained with concentrated colloidal suspension entailing a bulk frictional skeleton. **d)** Subharmonic peak $\omega_{1/2}$ and corresponding surface image for different $\Gamma$ in β-aescin solutions at $c \approx cmc$ (the different FW-states were tested; from II to V). The hydrodynamic crystals in the pure-FW state show the maximum structural stability characterized by the minimum peak broadening (at $\Gamma = 0.8$; magenta). **e)** Full width at half-maximum of the subharmonic peak ($\Delta\omega_{1/2}$) as a function of $\Gamma$ for pure water surface (black dots) and A/W-interfaces functionalized with aescin at 0.1 mM and 0.5 mM, respectively (red and blue dots). The minimum values determine the maximal crystal ordering. **f)** Minimum value of $\Delta\omega_{1/2}$ in e) as a function of aescin concentration, reported in terms of $G$. The solid line with the expected linear dependence between crystal order and shear rigidity represents a guide to the eye; for higher $G$ the minimum of $\Delta\omega_{1/2}$ decreases, which corresponds to a more stable and ordered hydrodynamic crystal upon stiffening the surface skeleton with increasing aescin concentration.

Furthermore, LDV-spectral analysis was exploited to quantify the intrinsic degree of hydrodynamic ordering achieved by the aescin-supported hydrodynamic crystals. We focused on the structural peak of subharmonic resonance ($\omega_{1/2}$), which determined the size of the unit cell ($\lambda_{1/2}$) (see Fig. 3a; bottom panels). The peak broadening $\Delta\omega_{1/2}$ can be interpreted as a signature of hydrodynamic ordering ($\Delta\omega_{1/2} \to 0$ for ideal hydrodynamic crystals). Figure 5d shows $\Delta\omega_{1/2}$ observed for increasing $\Gamma$ in aescin-supported FW-patterns. At the beginning of the pure FW regime, where incipient crystalline ordering began to emerge (Fig. 5d; right images), a finite broadening ($\Delta\omega_{1/2} > 0$ at $\Gamma \approx 2\Gamma_F$) was detected. The sharpest peak corresponded to the maximal degree of crystalline ordering (ca. $\Gamma \approx 4\Gamma_F$), still within the pure-FW domain. Further crystal distortion elicited progressive peak broadening ($\Delta\omega_{1/2} \gg 0$ at $\Gamma \approx 5\Gamma_F$), which was followed by hydrodynamic disordering until macroscopic melting in the chaotic regime ($\Delta\omega_{1/2} \to \infty$ at $\Gamma \approx 6\Gamma_F > \Gamma_{chaos}$). Figure 5e evidences the incremental effect of surface rigidization upon the intrinsic hydrodynamic ordering. Whereas a monotonic increase of $\Delta\omega_{1/2}$ with $\Gamma$ was observed for the bare water surface, increasing β-aescin concentration caused progressive peak narrowing within the pure FW interval. Remarkably, the most ordered crystals persisted to stronger excitations at higher aescin concentration (see caption in Fig. 5e). The material relationship is quantitatively evidenced in Figure 5f, which points out a phenomenological law $\Delta\omega_{1/2} \sim G^{-1}$. Analogous to the linear trade-off between shear modulus, density of defects and Bragg broadening existing in solids[39], our finding underlies the director role of surface rigidity in promoting the order in hydrodynamic crystals.

## Discussion

We proved the existence of free-standing hydrodynamic 2D-crystals as macroscopic patterns of unbounded FWs reversibly created as coherent excitations in water surfaces functionalized with a stiffening agent. From a synthetic standpoint, these hydrodynamic crystals can be moulded as FW-patterns in terms of two control parameters: excitation frequency and wave amplitude. Their crystalline structure was determined by the surface characteristics, which stem, exclusively, from the NLSW dispersion regime and from the high surface shear rigidity. By contrast to the conventional FW-patterns sustained upon bulk friction[40], and the container-dependent patterns of bounded waves[13,14,16], the new class of hydrodynamic crystals appeared under bulk material conditions that, without surface functionalization, would not allow the long-range order and the large crystal size achieved in our synthesis. The degree of crystalline order was determined by the shear stiffness: as a rule of thumb the higher G the lower the number of defects that distort the structure.

Once demonstrated the nature of the two-dimensional elastic stresses that "freeze" the FW-pattern leading to unbounded standing waves, a fundamental question came to mind: *What does constitute the hydrodynamic skeleton that is required to organize such an ordered FW-patterning?* Our answer was immediate from the evidences raised: *crucially, it is the mode coherence that instates disordered turbulence into flow organization.* Since temporal coherence enabled a hydrodynamic scaffold within the considered turbulent FW-cascades[41], it must persist stationary for a time longer than the fast energy exchange between the constituting modes[20]. Our experiments revealed crystal formation only upon coherent FW-turbulence exhibiting strong phase-locking between harmonic and subharmonic cascades materially harnessed by a skeleton of surface shear stresses enabled by a rigid adsorption monolayer. Therefore, the FW-supported hydrodynamic crystals here discovered consisted of both matter and waves, the former being guided by the later like in the de Broglie's picture of matter waves[42–44].

As a material condition, the adsorption monolayer of soluble β-aescin was revealed as necessary to endow enough stiffness for harnessing coherent FWs frozen in the surface plane. Aescin belongs to the group of saponins, which are biocompatible surfactants derived from glicoterpenoid compounds that endow optimal amphiphilicity due to a high packing efficiency[45]. Other surfactants that like β-aescin form rigid monolayers by reversible adsorption from aqueous media[21,22], could be also used to this purpose. The molecular ingredient of our hydrodynamic crystals adsorbs spontaneously covering the whole surface, thus β-aescin, or similar relatives, should be efficient in harnessing hydrodynamic crystals in all scales from macro to nano. Since the rigid monolayers of these (bio)surfactants enable interactions with biological membranes and self-assembling in solution[28], a plethora of unprecedented synthetic opportunities is envisioned in moulding artificial materials using hydrodynamic crystals.

In a panoramic view, our discovery challenges the classical intuition regarding the ability of liquid waves to mimic solid-like structures without the need of any previous phase transition and practically on demand. Undoubtedly, the finding of hydrodynamic 2D-crystals based on a completely liquid phase of matter provides a novel counterintuitive phenomenon that opens new possibilities to rethink the role of surface waves to manipulate matter.

## Outlook

The novel hydrodynamic crystals could be immediately translated into engineering and biological applications requesting structural encoding via a guiding wave field. The moulding versatility under external control makes our method meaningful for rescaled implementation in substrate-guided approaches to the synthesis of macro- to nano-structured (bio)materials[47,48]. Boosted by nanotechnological requests to mimic condensed-matter systems[49,50], our straightforward FW-platform could embed functional objects into ordered lattices with an externally tunable size and symmetry. Furthermore, the study of topological defects in hydrodynamic liquid crystals will open the possibility to study the formation of currents to drive movements and directional transport[51]. In a more fundamental perspective, FW-patterns could result into a wave–crystal association conceptually linked to the de Broglie's duality for extended systems in two dimensions[42,52]. Indeed, FW's have been argued as the classical paradigm of de Broglie duality[42], in which the physical nature of the guiding FW-field has been unequivocally stablished[44,53]. As a wet-lab realization of the Feynman's simulator[54], our hydrodynamic crystals could result useful to manipulate 2D-matter waves at interaction with different materials and external fields.

To summarize, our elegant "hydrodynamic crystals" not only constitute an innovation for future developments in material synthesis with soft and biological matter but also a novel archetype of waving ordering in two dimensions, which could shed light on the interactions between matter and waves at the critical crossover from the weak hydrodynamic turbulence to the Faraday regimes where wave order emerges as a crystal in two dimensions.

## Materials and methods

Ultrapure water was from a Milli-Q source (Millipore). All the chemicals and surfactants used, including β-aescin, were from Sigma. For improving surface visualization, highly diluted aqueous suspensions of PS-MAA and PS microparticle were synthetized by surfactant-free emulsion polymerization (see Supplementary Note N1).

**Parametric NLSW excitation as gravity-capillary (GC) waves.** Liquid surfaces under external forcing at a single frequency $\omega_0$ respond as an ensemble of nonlinear surface waves (NLSW) of master wavelength $\lambda_0$, which defines the fundamental mode determined by a natural frequency equalized to the forcing frequency $\omega_0$. In the linear regime (Hookean), only the fundamental response appears as a monochromatic GC-wave restored by gravity and surface tension. In the deep-water approximation, the GC-dispersion relationship is given as $\omega^2 = gk + \sigma k^3/\rho$ where $g$ is the acceleration of gravity and $k = 2\pi/\lambda$ the wavevector; $\sigma$ and $\rho$ accounts, respectively, for the liquid surface tension and density[55]. At frequencies lower than the capillary frequency $\omega_c = (\rho g^3/\sigma)^{1/4}$, one founds G-waves of large wavelength controlled by the acceleration of gravity, this is $\lambda_g \approx 2\pi g/\omega^2$ at $\omega < \omega_c$; in water, $\omega_c \approx 14\ Hz$ and $\lambda_g \geq 1\ cm$. At higher frequencies above the capillary frequency ($\omega > \omega_c$), a crossover occurs to the capillary regime dominated by surface tension; here, the C-waves have a shorter wavelength ($\lambda_c \ll 7\ mm$) varying with frequency as $\lambda_c \approx 2\pi(\sigma/\rho\omega^2)^{1/3}$. The presence of in-plane rigidity due to a viscoelastic film causes the capillary waves to propagate at a slower velocity characterized by smaller wavelengths than in the bare surface[56]. We have deduced a perturbative expansion for the elasticity-dependent GCW-dispersion relation in the relevant limit of dominating rigidity, i.e. at $G/\sigma \gg 1$ (see Supplementary Note N2). Nonlinear GC-waves can be also excited in different NLSW-domains[30], which can be tuned as a parametric excitation in terms of the driving amplitude $(A)$[22]. At low amplitudes below a Hookean threshold $(A < A_0)$, only the linear response corresponding to the fundamental GC-mode excited at $\omega = \omega_0$ is found. At high amplitudes $(A > A_0)$, GC-NLSWs emerge as cascades of resonant harmonics of the fundamental GC-response ($\omega_n = n\omega_0$ with $n = 2,3,...$)[57]. Energy and symmetry conservation impose all the GC-harmonics with a same wave dispersion than the fundamental driving mode[30].

**Surface wave exciter.** We employed a cylindrical vessel fabricated in Plexiglas® with a design optimized to minimize external vibration and meniscus effects. In order to assure waving patterns with a large aspect-ratio, we designed the vessel with dimensions large enough with respect to the typical capillary wavelengths (diameter $D = 20\ cm$; height $h = 2\ cm$, thus $D, h \gg \lambda \leq 2\ mm$). Vertical vibrations of frequency $\omega_0$ are induced by a louder connected to a power amplifier driven by a function generator (Agilent 33220A). A band-pass filter (Stanford SR560) is used to prevent for spurious harmonics in the driving signal; this guarantees monochromatic excitation as $z(t) = A cos(\omega_0 t)$, where $A$ is the surface vertical displacement of the container. GC-waves are spatially damped along a characteristic length that depends on the ratio of the surface stresses and the bulk friction. The largest decay length is determined by the longest gravity mode, i.e. $l_g \approx \rho g(\lambda/2)A/4\omega\eta \approx g^2 A/4\mu\omega^3$; in water (with a kinematic viscosity $\mu \approx 2\ 10^{-6}\ m^2/s$), for the largest vertical amplitudes ($A \leq 2\ mm$), elicited at the slowest driving frequencies (ca. 10 Hz), one estimates

$l_g \leq 10\ cm$, which fixes an adequate vessel diameter at $D = 20\ cm \gg l_g$. The oscillatory liquid displacements drive the surface against restoring gravity and capillary forces at acceleration uniquely determined by the excitation characteristics as $a = A\omega_0^2$. Only a limited $a$-range was experimentally accessible below the capillary frequency, since for $\omega_0 \ll \omega_c$, one has $a \ll g$. In order to avoid vibration parasites, the experiments were placed on an antivibration table (Newport).

**Laser Doppler Vibrometry (LDV).** Surface vertical displacements were detected by a laser Doppler vibrometer (Polytec PDV100). A HeNe laser beam (632.8nm; 1 mW) is focused into a surface spot placed in the center of the liquid vessel. The reflected signal is analyzed in an interferometric detection scheme to retrieve the surface normal velocity in the time domain. The power spectral density ($PSD$) is determined by Fourier transformation of the time series of surface velocities acquired by LDV. This setup allows us to perform a simultaneous investigation over a large spectral range covering frequencies up to $2\ 10^4\ Hz$. The vertical acceleration ($a$) was determined as the derivative of the surface velocities measured by LDV. Negligible water evaporation, neither changes in surface tension nor temperature drifts occurred during the experiments. Optical imaging was performed with a CCD camera (Nikon D5600) under white light illumination of the water surface. To enhance optical contrast a few droplets of PS-MMA suspension were dissolved till milky shadowing of the liquid surface.

**Interfacial rheology.** We employed a hybrid rheometer (TA instruments, DHR20) working in the interfacial mode with a ring tool oscillating at variable frequency ranging 0.1-100 Hz and a fixed 0.1% shear strain in which the surface response is completely linear (see Supplementary Note N3).

## Acknowledgements

We thank M.G. Velarde and J. Santamaria for fruitful discussions. The authors acknowledge Comunidad de Madrid for funding this research under grants Y2018/BIO-5207 and S2018/NMT-4389, and Ministerio de Ciencia e Innovación (MICINN) under grant PID2019-108391RB-I00. We thank our colleague Dr. E. Montoya, who provided insight and expertise with the LDV device, and Prof. J. Fernández-Castillo for generosity in free leasing laboratory space.

## Author contributions

FM designed the research, provided the experimental methods and discussed the results. MK performed the experiments, analyzed the data and discussed the results. NC and HLM analyzed the data and discussed the results. EE provided materials and discussed the results. MK, NC and FM wrote the paper.

**Affiliations:**

1. **Department of Physical Chemistry. Complutense University of Madrid. Spain.** Mikheil Kharbedia, Niccolo Caselli, Horacio López-Menéndez, Eduardo Enciso and Francisco Monroy.

2. **Unit of Translational Biophysics. Institute for Biomedical Research Hospital Doce de Octubre. Spain.** Francisco Monroy.

## Competing interests

The authors declare no competing interests.

# Supplementary Material

## Moulding hydrodynamic 2D-crystals upon parametric Faraday waves in shear-functionalized water surfaces.

Mikheil Kharbedia[1], Niccolò Caselli[1], Horacio López-Menéndez[1], Eduardo Enciso[1] and Francisco Monroy[1,2,*]

[1]Department of Physical Chemistry, Universidad Complutense de Madrid, Ciudad Universitaria s/n E28040 Madrid (Spain).
[2]Unit of Translational Biophysics, Instituto de Investigación Sanitaria Hospital Doce de Octubre, E28041 Madrid (Spain).
**e-mail:** monroy@quim.ucm.es

## Table of Contents

**Supplementary Tables T1-T2**

**Supplementary Figures S1-S10**

**Supplementary Notes N1-N3**

## Supplementary Tables T1-T2.

**Supplementary Table T1. Critical acceleration thresholds for the non-linear surface-wave (NLSW)-states excited on the free surface of different liquids.** We considered bare water surface with a high surface tension, low-tension aqueous surface with a CTAB-adsorption layer *(light gray)*, and high-viscosity water/glycerol mixture *(dark gray)*. We report the experimental values of critical acceleration ($a_X$, where X stands for L: linear; KZ: Kolmogorov-Zakharov; F: Faraday; C: Chaotic) on the onset of different parametric wave states as quantitatively identified by the slope of the turbulence cascade detected by means of laser Doppler vibrometry (LDV). The onset for Faraday instability ($a_F$; marked in blue) measured as a function of the driving frequency ($\omega_0$) follows the theoretical prediction as given by Eq. (1) of the main text for the quoted values of surface tension (σ) and the kinematic viscosity (η) (see Supplementary Figure S1 and Fig. 3c and the main text for details). The Faraday wave (FW)-dispersion status is determined by the value of $\omega_0$ with respect to the capillary frequency $\omega_c$ (for water, $\omega_c = 14\ Hz$). Thus, gravity waves occur below or close or below $\omega_c$, whilst capillary waves emerge for $\omega_0 > \omega_c$. In order to study the FWs in a defined dispersion regime, the experiments reported in the main manuscript corresponded to the reference frequency $\omega_{ref} = 47 Hz$, i.e. in the capillary regime. By analysing the results reported in Table T1, the dominance of inertia in creating NLSWs is evident, since stronger forces are needed with decreasing surface tension (in CTAB solutions) and increasing viscosity (in water/glycerol mixtures).

.

| Liquid surface | $\omega_0$ (Hz) | σ (mN/m) | η (mPa.s) | $a_L$ (m/s$_2$) | $a_{KZ}$ (m/s$_2$) | $a_F$ (m/s$_2$) | $a_C$ (m/s$_2$) |
|---|---|---|---|---|---|---|---|
| water ($\omega_0 < \omega_c$) | 10 | 72 | 1 | 0,0001 | 0,0005 | 0,001 | 0,003 |
| water ($\omega_0 \approx \omega_c$) | 15 | 72 | 1 | 0,0002 | 0,0006 | 0,002 | 0,006 |
| water ($\omega_0 > \omega_c$) | 25 | 72 | 1 | 0,0004 | 0,001 | 0,006 | 0,03 |
| water ($\omega_0 \gg \omega_c$) | 35 | 72 | 1 | 0,001 | 0,03 | 0,01 | 0,6 |
| water ($\omega_0 = \omega_{ref}$) | 47 | 72 | 1 | 0.01 | 0.8 | 2 | 10 |
| water + 0,1mM CTAB | 47 | 62 | 1 | 0.02 | 1.2 | 4.1 | 22 |
| water + 1mM CTAB | 47 | 55 | 1 | 0.02 | 1.9 | 6.7 | 25 |
| water ($\omega_0 > \omega_{ref}$) | 60 | 72 | 1 | 0.01 | 1.1 | 2.3 | 15 |
| water ($\omega_0 \gg \omega_{ref}$) | 73 | 72 | 1 | 0.01 | 1.6 | 3 | 25 |
| water + 5% glycerol | 47 | 72 | 6 | 0.2 | 1 | 2 | >30 |
| water + 15% glycerol | 47 | 72 | 28 | 0.2 | 2.3 | 9 | >30 |

**Supplementary Table T2. Critical acceleration thresholds for parametric surface wave states in water covered with aescin.** To study the appearance of NLSWs in the different regimes of wave propagation, we focused on the amplitude transitions at variable acceleration in three well-delimited dispersion regimes; these are: hybrid gravity-capillary waves close to the capillary frequency (GCW at $\omega \approx \omega_c = 14 Hz$), pure gravity waves (GW at $\omega \ll \omega_c$) and pure capillary waves (CW at $\omega \gg \omega_c$). We detected the acceleration thresholds as reported in Table T1 (the onset of the Faraday instability is marked in blue). No significant differences were detected in comparison with the bare surface of water (see Supplementary Table T1).

| $\omega_0$ (Hz) | dispersion | $\sigma$ (mN/m) | $\eta$ (mPa.s) | $a_L$ (m/s$_2$) | $a_{KZ}$ (m/s$_2$) | $a_F$ (m/s$_2$) | $a_c$ (m/s$_2$) |
|---|---|---|---|---|---|---|---|
| 10 | GW | 40 | 1 | 0,0001 | 0,0003 | 0,001 | 0,005 |
| 15 | CGW | 40 | 1 | 0,0002 | 0,005 | 0,02 | 0,04 |
| 25 | CGW | 40 | 1 | 0,005 | 0,01 | 0,04 | 0,07 |
| 35 | CW | 40 | 1 | 0.01 | 0.16 | 0.5 | 0,6 |
| 47 | CW | 40 | 1 | 0.01 | 0.4 | 1.6 | 9 |
| 60 | CW | 40 | 1 | 0.05 | 0.5 | 3 | 14 |
| 73 | CW | 40 | 1 | 0.1 | 1 | 5 | 25 |

# Supplementary Figures S1-S9

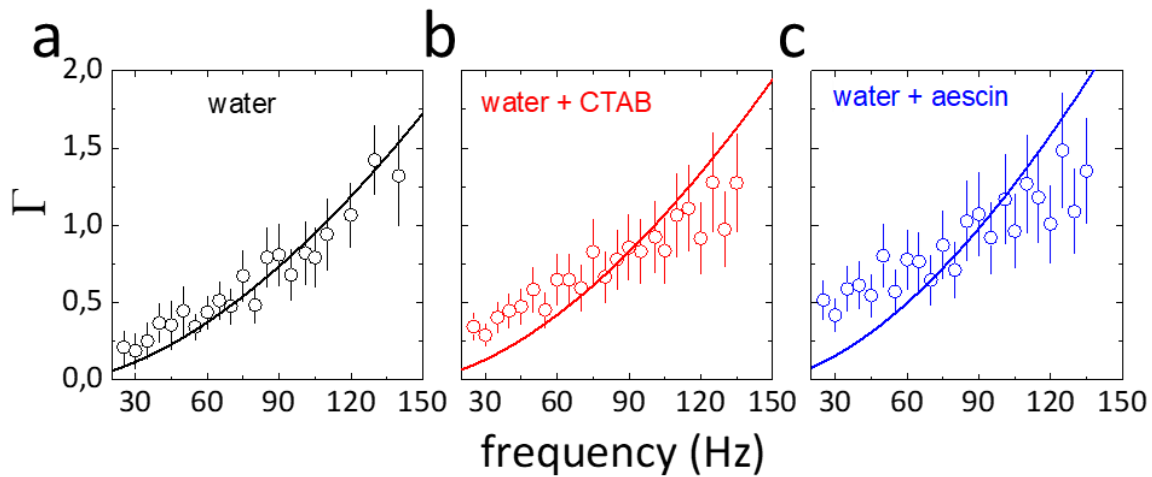

**Supplementary Figure S1. Onset of the Faraday instability for the bare water surface and functionalized with adsorbed monolayers.** The critical acceleration $\Gamma = a/g$ was measured as a function of excitation frequency for Faraday waves in pure water (a), and in aqueous solutions of CTAB 1.5 mM (b) and aescin 1.0 mM (c). Both concentrations are above the respective cmc. The circles correspond to experimental data determined as the onset of the Faraday instability observed at increasing excitation amplitude (the average values correspond to at least five independent determinations with standard deviation specified as error bars). The solid curves correspond to Eq. (1) of the main manuscript, which reproduces the expectation for the critical acceleration to induce Faraday waves restored by surface tension (water: $\sigma = 72\,mN/m$, CTAB: $\sigma = 50\,mN/m$, aescin: $\sigma = 30\,mN/m$,) in a bulky liquid enclosed by a vibrating receptacle ($\rho = 10^{-3}\,kg/m^3$ and $\eta = 10^{-3}\,Pa\,s$). Note that the data agree with theoretical predictions, although in the case of water surface covered by aescin a small deviation at low frequency is found.

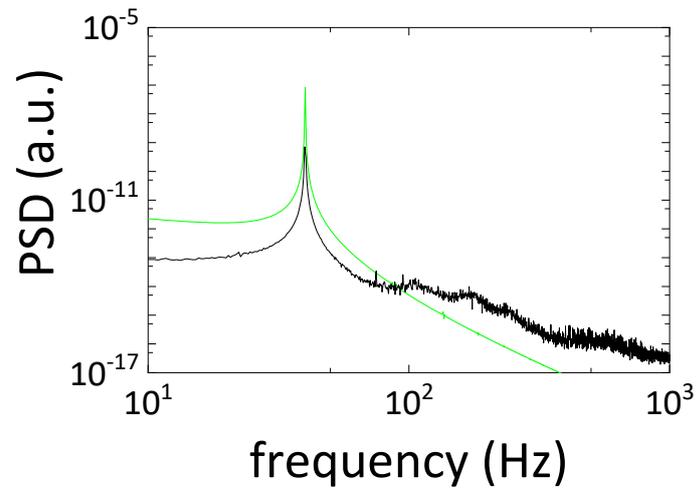

**Supplementary Figure S2. Linear response at low driving amplitudes.** Surface wave spectrum in the linear response regime (black curve), obtained for very low vertical excitation amplitude $\Gamma = 0.03 \ll \Gamma_F$, at $\omega_0 = 47$ Hz (capillary regime). The observed resonance occurs exactly at the pumping frequency induced by the lauder (green curve). The observed spectral broadening is compatible with the natural bandwidth of the excitation channel (green straight line).

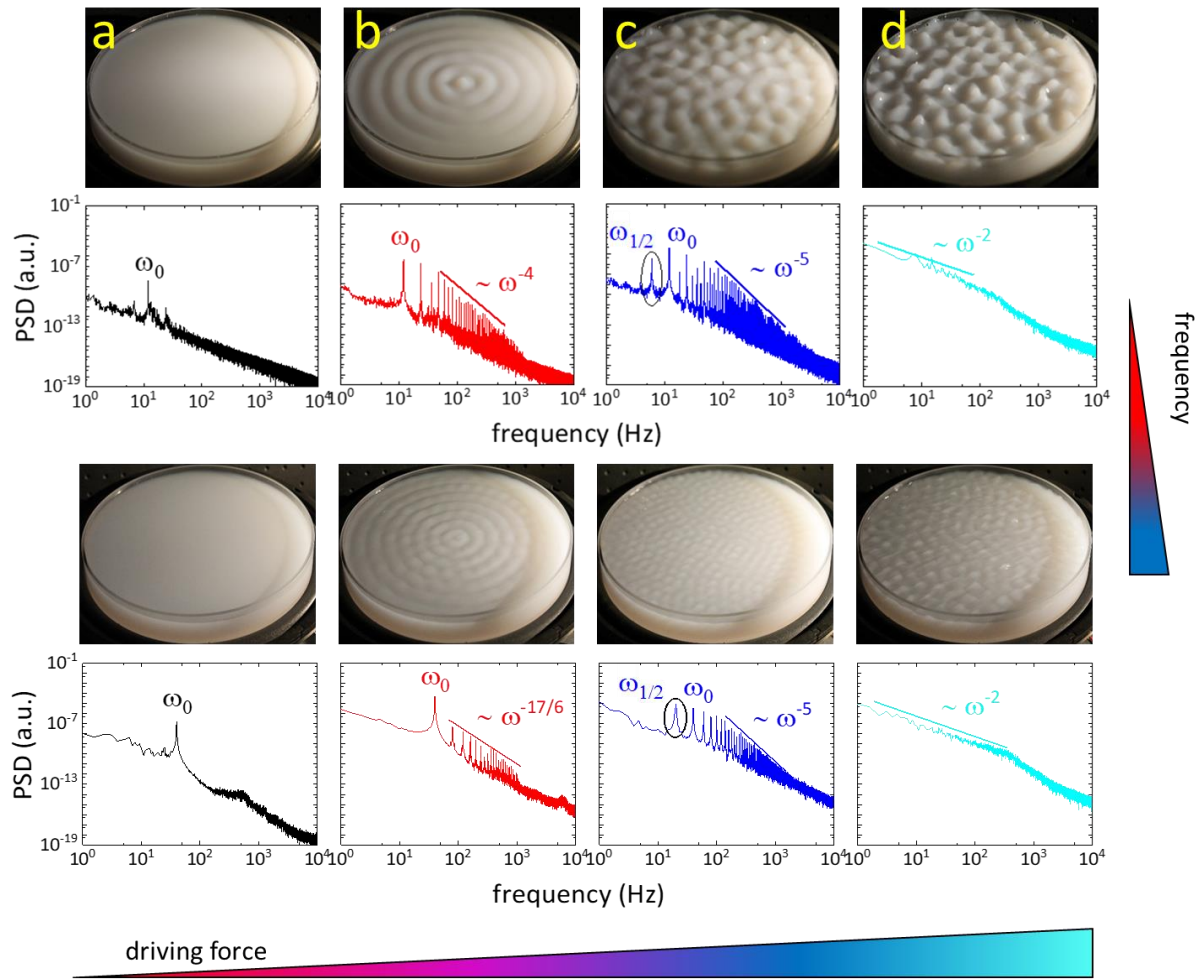

**Supplementary Figure S3. NLSWs regimes in water surface.** Images and spectral response of nonlinear surface waves driven on pure water under parametric excitation ($\omega_0 = 12 Hz$ and $\omega_0 = 47 Hz$, upper and down panels, respectively). For increasing vertical acceleration determining the driving force (from a to d) different wave regimes were observed, as described in the main paper: (a) linear regime exhibiting a single peak at $\omega_0$, (b) Kolmogorov-Zakharov spectrum of weak wave turbulence exhibiting container-shape surface waves and a $\omega^{-17/6}$-intensity decay typical of cascades of nonlinear capillary waves, (c) Faraday regime exhibiting disordered surface roughening and a spectrum with subharmonic response at $\omega_0/2$, and an intensity decay cascade that scales as $\omega^{-5}$; (d) Chaotic regime featuring strongly disordered surface and a continuous spectrum (without any particular resonance) and a intensity decay proportional to $\omega^{-2}$.

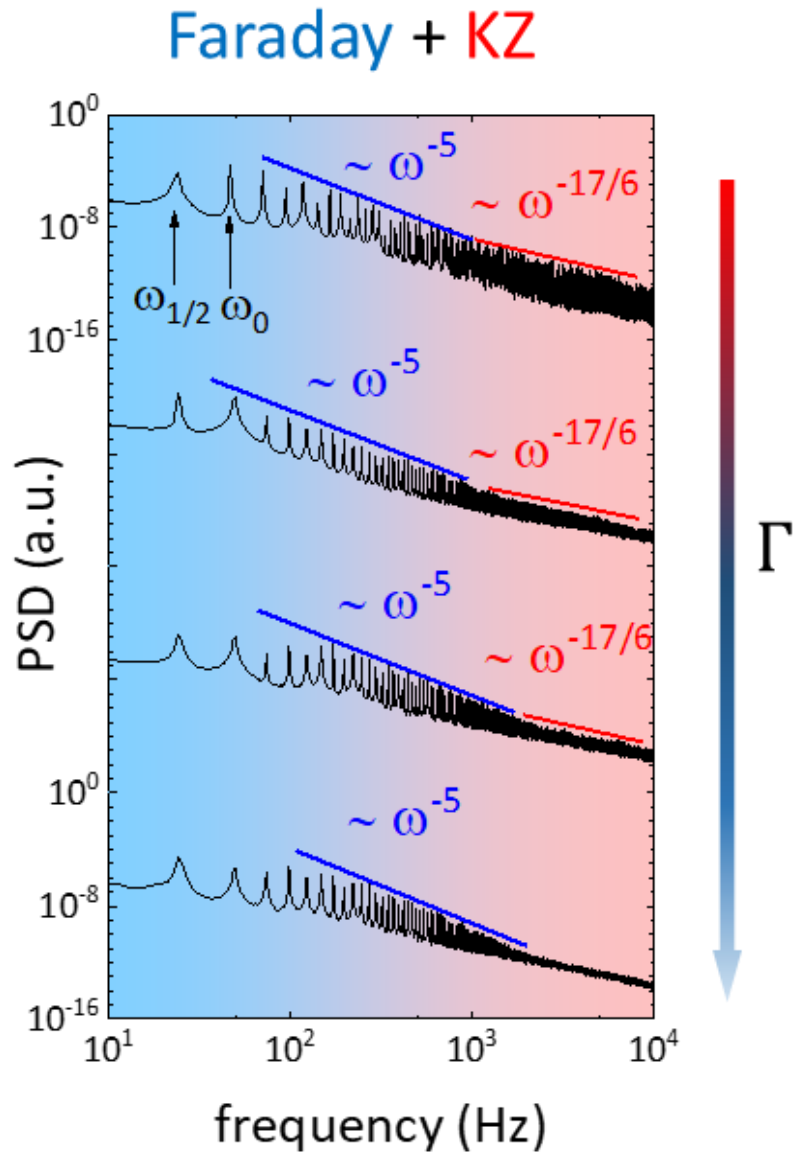

**Supplementary Figure S4.** Hybrid spectrum of turbulence for pure water in the capillary regime ($\omega_0 = 47 Hz$) where both Faraday ($\sim \omega^{-5}$) and Kolmogorov-Zakharov (KZ, $\sim \omega^{-17/6}$) spectral decay coexist. The former was found in the low frequency range, while the latter in the high frequency spectrum. For increasing $\Gamma$ ($0,07 - 2,5$) the hybrid spectrum converts into a pure Faraday wave cascade. As discussed in the main manuscript, the KZ spectrum is restricted to the higher frequencies range due to its lower intensity in comparison with Faraday wave cascade.

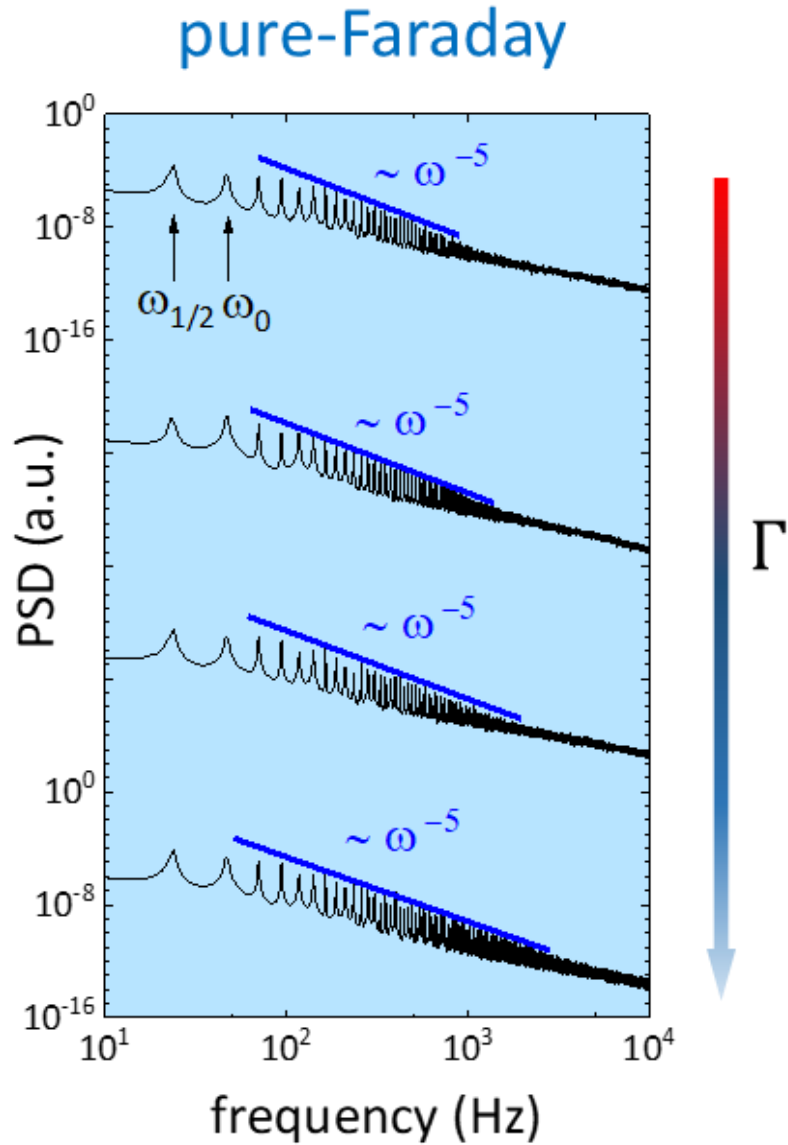

**Supplementary Figure S5.** Pure Faraday wave spectra (with intensity decay $\sim \omega^{-5}$) in the capillary regime ($\omega_0 = 47 Hz$) for bare water surface, depicted in the $\Gamma$ range $(0,3 - 0,7)$ to avoid hybrid spectrum of turbulence.

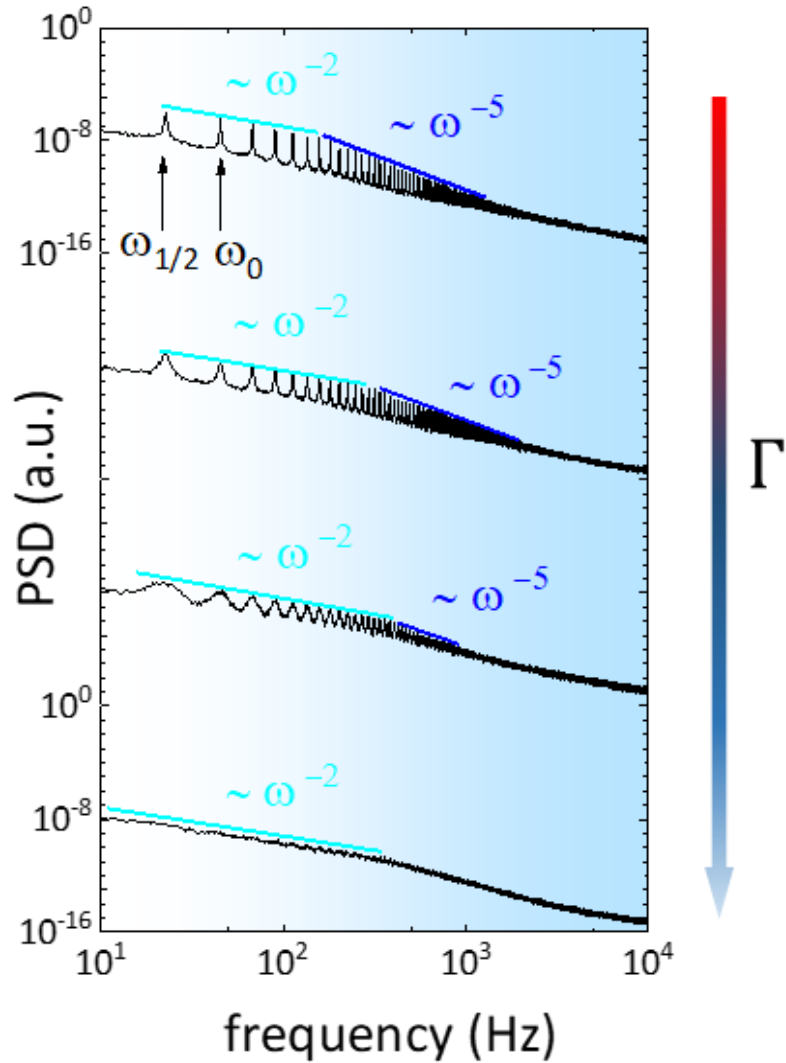

**Supplementary Figure S6.** Hybrid spectrum of turbulence obtained at the transition from Faraday to continuous spectrum for bare water surface in the capillary regime ($\omega_0 = 47 Hz$) by increasing $\Gamma$ in the range $(0{,}8 - 2)$. At higher values of $\Gamma$ continuous spectrum emerges with characteristic Brownian-like energy cascade ($\sim \omega^{-2}$). This cascade firstly develops in the lower frequency range and, for increasing $\Gamma$, propagates to higher harmonics. Also, it is noteworthy the fact that for increasing excitation amplitudes the spectral peak broad until they disappear at higher $\Gamma$.

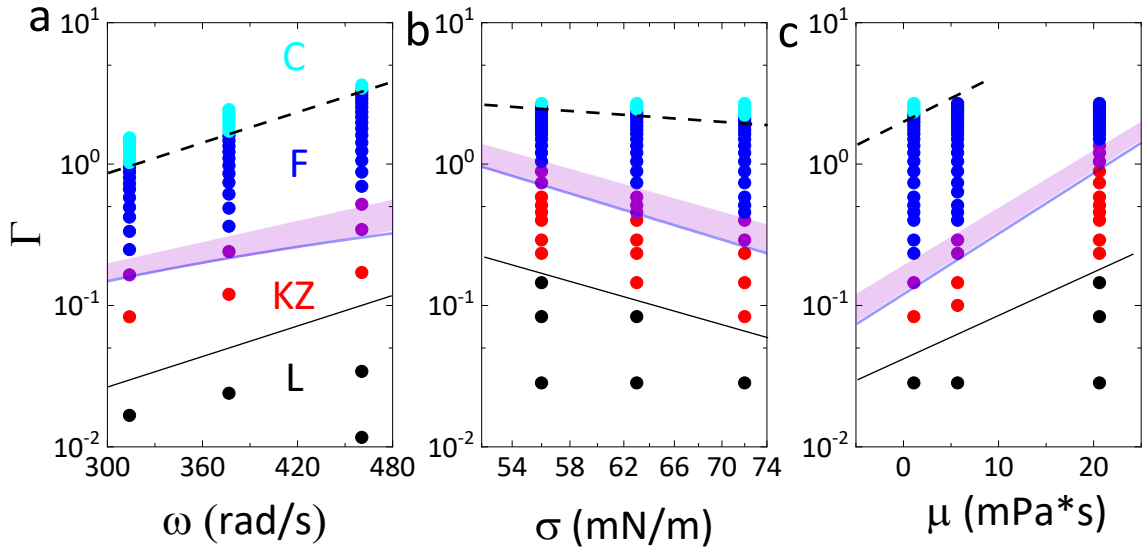

**Supplementary Figure S7. NLSW-states in liquid surfaces of variable surface tension and viscosity. a)** State diagram of nonlinear surface waves in water under parametric excitation, as a function of excitation frequency $\omega_0$ and the reduced acceleration $\Gamma$. **b)** State diagram for different surface tension $\sigma$, varied by increasing CTAB concentration close to its cmc. **c)** State diagram for different liquid viscosity $\mu$, changed by increasing the glycerol concentration in water. In each of the three presented scenarios, five different states are observed: i.e. linear (black dots), KZ (red dots), hybrid KZ-Faraday (pink band), pure Faraday (blue dots) and continuous spectrum (light blue dots). The blue solid line corresponds to Eq. (1) of the main manuscript, which coincides with the KZ to Faraday spectrum transition. The black dashed line indicates the experimental transition from Faraday to continuous spectrum. The black solid line the transition from linear to KZ spectrum. Finally, the pink area represents the hybrid F-KZ spectral region.

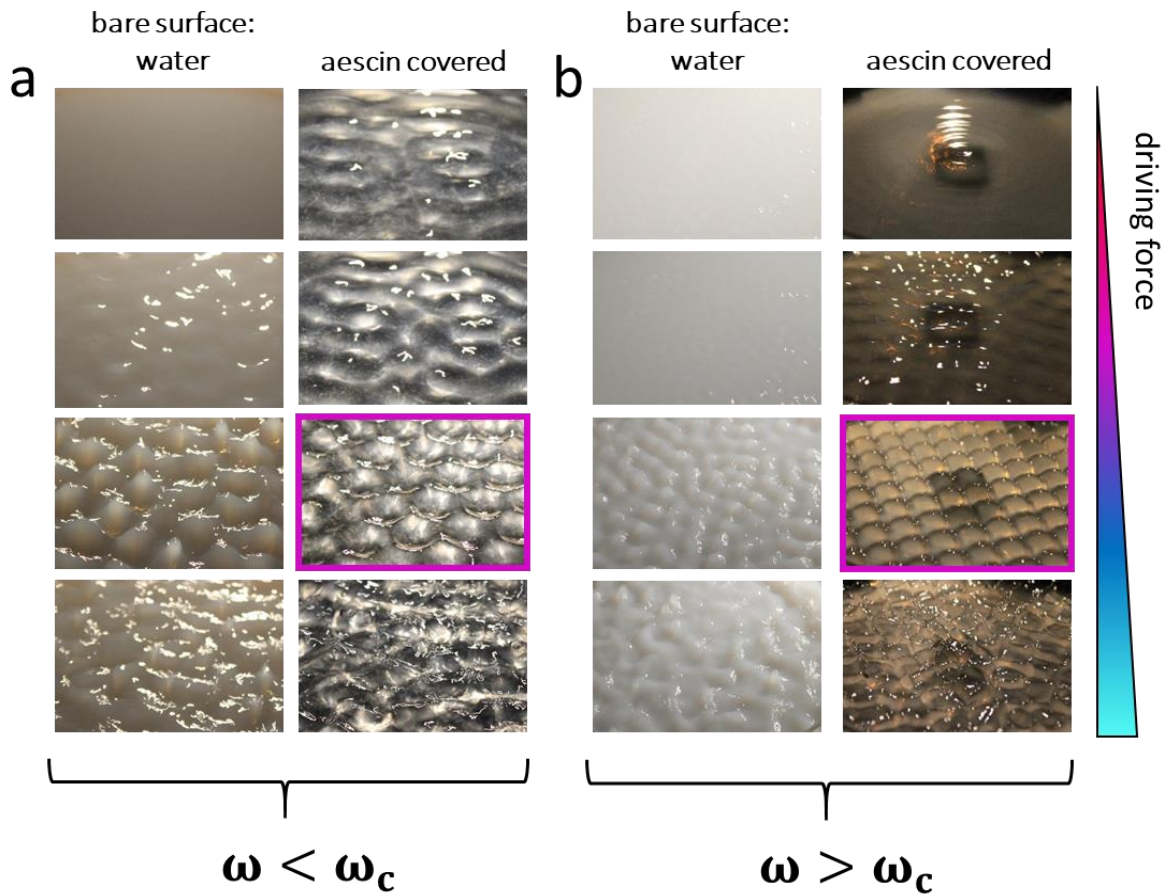

**Supplementary Figure S8. a)** Comparison of images of bare water (left panels) and aescin covered (right panels) surface waves driven in the gravity regime ($\omega_0 = 10 Hz$) for increasing vertical acceleration (from top to bottom panels). The aescin concentration was 1 mM. Only the introduction of aescin was able to induce an ordered surface pattern on the Faraday waves (magenta surrounded box) **b)** Comparison of images of bare water (left panels) and aescin covered (right panels) surface waves driven in the in the capillary regime ($\omega_0 = 47 Hz$) by increasing the vessel vertical acceleration (from top to bottom panels). By increasing the vertical acceleration, surfaces waves emerged due to the parametric resonance and organized in a regular packing leading to the formation of a hydrodynamic crystals which achieved their maximum packing order at intermediate accelerations (magenta surrounded boxes). By further increasing the driving acceleration a crystal breakup occurred.

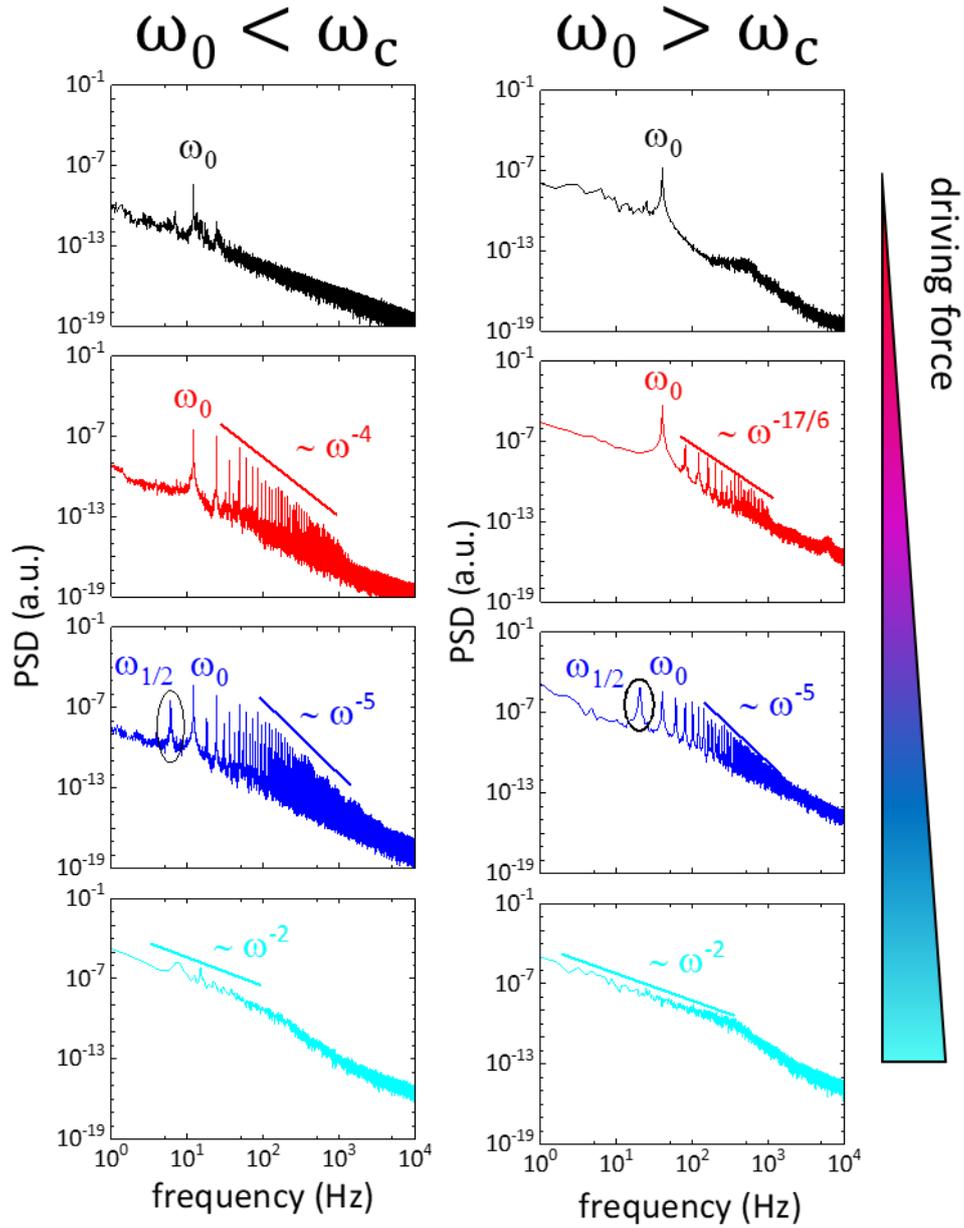

**Supplementary Figure S9.** Representative log-log spectra of NLSW turbulence in aescin covered water surface. Below the capillary frequency one finds GWs (left panels), and CWs above of (right panels)). The sequence of spectra spans from top with the linear regime (only the fundamental response is present at $\omega_0$) to bottom; then, the incoherent KZ-turbulence characterized by a spectral decay exponent $\alpha$ that depends on the dispersion regime ($\alpha = -4$ for GWs and $\alpha = -17/6$ for CWs), through of the Faraday instability (where the subharmonic response appears at $\omega_{1/2} = \omega_0/2$ and the FW-turbulent cascades decay as $\alpha = -5$), up to the chaotic regime in the lower panels ($\alpha = -2$). Crystal formation is only observed in the Faraday regime ($\alpha = -5$) (see main text for details).

## Supplementary Notes N1-N3.

**Supplementary Note N1. Synthesis of aqueous suspensions of PS-MAA and PS microparticles.**

*PS-MAA particle suspension.*

*Synthesis.* Aqueous suspensions of spherical particles of Poly(styrene-co-methacrylic acid) (PS-MAA) were prepared from surfactant-free emulsion by a conventional polymerization process of both monomers [M. C. Carbajo, E. Climent, E. Enciso and M.J. Torralvo, "Characterization of latex particles arrays by gas adsorption", J. Coll. Interface Sci., 284 (2005), 639-645]. The polymerization reaction was carried out in a 500 mL round-bottomed, five-necked flask. In the flask outlets, a water-cooled reflux condenser, a T-shaped stirrer, a gas inlet and a contact thermometer were fitted. One outlet was used for introducing the chemicals of the reaction. The flask was introduced in a thermostat water bath which controls the reaction temperature to $\pm 2°C$. The synthetic process starts by introducing 253 mL of water, 18 g of styrene and 1.58 g of methacrylic acid. The styrene was washed with a NaOH 0.1M water solution to remove the monomer inhibitors. The stirring was fitted to 350 rpm and nitrogen gas was bubble during all reaction time to remove any oxygen traces. When the temperature arrived to 85 ºC, 0,22 g of potassium persulfate (KPS) was added as initiator of the polymerization reaction. After eight hours of reaction, the flask is removed from the bath and cooled in a water-ice mixture. The resulting polymer suspension is dialyzed against water over three weeks.

*Results.* Scanning electron microscopy (SEM) was performed with a JEOL-JSM-6330F electron microscope operating at 10kV. By analysing SEM-images using ImageJ, the average diameter was estimated into 190 nm. Measurements of dynamic light scattering (DLS) with a 90-Plus/BI-MAS apparatus was performed to characterize particle size distribution. An average particle diameter of 200 nm was obtained, with a relative standard deviation of 0.030 indicating an extremely low size polydispersity. Electrophoretic mobilities were estimated by using the ZETA-PALS mode of the BI-MAS equipment. The measures were performed at room temperature and an ionic strength of 1mM of $KNO_3$. The analysis of the data gave a zeta-potential of 47.3 mV on negative charged particles, with charges come from the persulfate fragments.

*PS particle suspension*

*Synthesis.* We also prepared Polystyrene (PS) microparticles by following the same synthetic route as with PS-MMA, without adding MAA monomer and introducing a small monomer charge of the reactor (9 g). The slow polymerization rate of styrene required to perform the polymerization during 24h.

*Results.* A dilute PS suspension was measured by DLS and the average particle diameter was 313 nm, with a relative standard deviation of 0,037.

**Supplementary Note N2. Dispersion relation of GC-waves in the presence of a surface viscoelastic film.** We are interested in the correction to GC-wave dispersion due to surface elasticity arising from a viscoelastic film adsorbed to the surface. The classical treatment of Hansen and Mann (HM) considers linear hydrodynamics for the propagation of capillary waves on viscoelastic surfaces [R. S. Hansen and J.A. Mann, J. Appl. Phys. 35(1), 152 (1964)]. To account for the case of surface waves propagating on aescin covered surfaces as the data represented in the Figure 3b of the main text, we considered the HM-theory for a highly rigid surface film, this is at the material limit $\xi = G/\sigma \gg 1$. Essentially, the viscoelastic film is known to affect capillary motions inducing weak changes on the propagation characteristics ($\omega$ and $k$) companioned by strong wave damping ($\beta$). For the propagation frequency of the capillary waves, the HM-theory predicts a non-monotonic variation with the capillary wavenumber, from the ideal Kelvin value corresponding to the bare surface, this is $\omega_0 = (\sigma k^3/\rho)^{1/2}$ for $\xi = 0$, through an intermediate maximum of optimal propagation rate that is dependent on the elasticity ratio $\xi = G/\sigma$, down to the limit $\xi \to \infty$ ($G \gg \sigma$), where the propagation frequency reaches a high rigidity value $\omega_\infty = \sigma k/\eta$, which is completely independent of $\xi$ but compatible with the overdamped solution for a capillary ripple in the bare surface of a liquid of viscosity $\eta$. Because aescin monolayers adsorbed at concentrations close to cmc are terribly rigid ($\xi \approx 30$, since $G \approx 1\, N/m$ and $\sigma \approx 0.03\, N/m$) as compared to typical adsorbed films of typical surfactants $\xi \ll 1$, we considered the HM-picture of a highly viscoelastic film to describe the propagation characteristics of the surface waves in the presence of a rigid monolayer of aescin adsorbed at the free surface of water.

Briefly, we first took the ideal solution for GC-waves considered in Methods, and then reorganized it with the reduced form considered in HM's work:

$$y(\omega) = \frac{\rho \omega^2}{\rho g k + \sigma k^3}$$

Because the perturbative corrections to the ideal solution $y(\omega_{GC}) = 1$ in the HM-theory can be expressed in terms of the elasticity ratio $\xi = G/\sigma$ and the dimensionless factor $u = \eta\omega/\sigma k$, further elaborating on the result for a highly viscoelastic film (Eq. 38 in the 1964's paper), in the limit $\xi \to \infty$ we obtained the approximate formula:

$$y_{\xi \to \infty}(\omega) \approx 1 - \left(\frac{\omega}{2\omega_\infty}\right)^{1/2} + \frac{1}{4}\frac{\omega}{\omega_\infty}$$

where we have substituted the factor $u = \eta\omega/\sigma k = \omega/\omega_\infty$ in the HM-theory as written in terms of the overdamped frequency $\omega_\infty = \sigma k/\eta$.

This equation represents the high-rigidity perturbation limit to the GCW propagation in a free surface covered with a viscoelastic film. It was numerically resolved to obtain the limiting solution $\lambda_{\xi \to \infty}(\omega)$ plotted in Fig. 3b of the main paper.

**Supplementary Note N3. Experimental setup for the interfacial rheology measurement of the in-plane shear modulus of the functionalized surfaces.** We used a commercial hybrid rheometer (HDR100, TA Instruments) working in the surface shear mode for measuring the viscoelastic modulus modulus of surface films at the air/water interface. **a)** Du Nöuy ring configuration with the thermostatic vessel. **b)** The ring is in contact with the aqueous solution surface with surfactant, which develops a surface monolayer (here simplified by drawing a schematic head-tail amphiphilic molecule configuration). **c)** The rheological measurement is carried out by in-plane shear force stressed on the surface monolayer by the ring tool. **d)** Experimental measurements of the stress-strain relationship in the same systems considered in this work. Data in the Figure 5 of the main text were measured at 0.1% strain, corresponding to the linear regime.

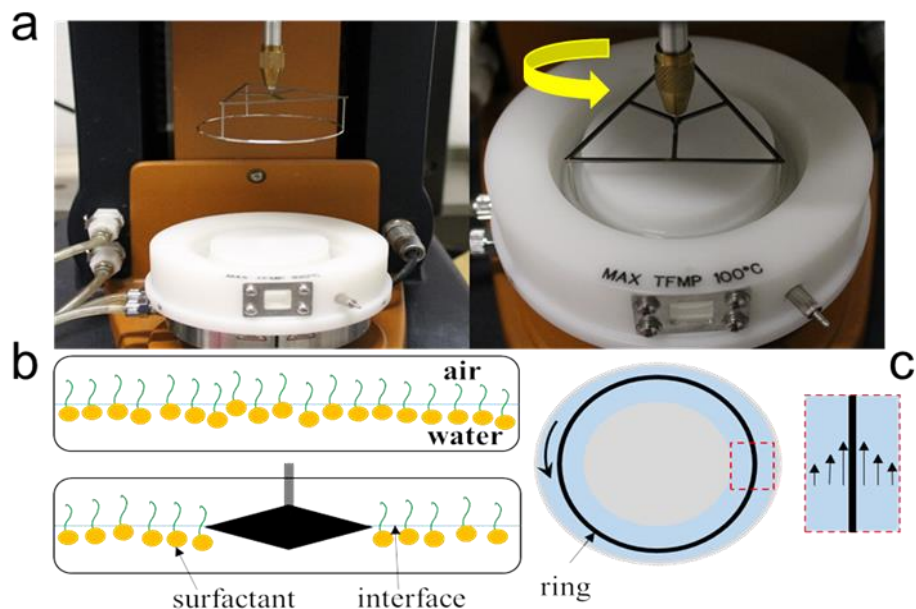

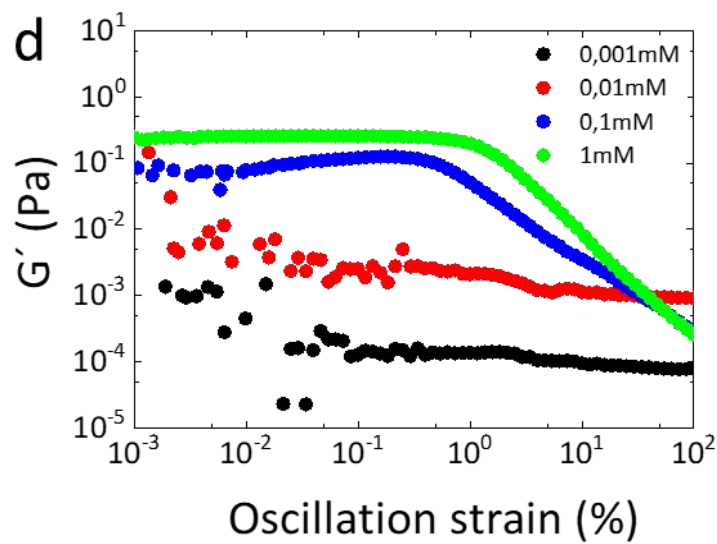